\documentclass[aps,prl,nofootinbib,twocolumn,superscriptaddress,floatfix,notitlepage]{revtex4-1}
\pdfoutput=1

\usepackage{graphicx}
\usepackage{amsmath,amsfonts,amssymb}
\usepackage{color}
\usepackage[breaklinks,colorlinks,urlcolor=blue,citecolor=blue,linkcolor=magenta]{hyperref}
\usepackage{verbatim}
\usepackage{enumitem}
\usepackage{hyperref}

\usepackage{aas_macros}

\newcommand{\td}{{\rm d}}
\newcommand{\vect}[1]{\boldsymbol{#1}}

\newcommand{\be}{\begin{equation}}
\newcommand{\ee}{\end{equation}}
\newcommand{\bea}{\begin{equation} \begin{aligned}}
\newcommand{\eea}{\end{aligned} \end{equation}}

\def\lsim{\mathrel{\raise.3ex\hbox{$<$\kern-.75em\lower1ex\hbox{$\sim$}}}}
\def\gsim{\mathrel{\raise.3ex\hbox{$>$\kern-.75em\lower1ex\hbox{$\sim$}}}}

\newcommand{\papertitle}{What is the origin of the JWST SMBHs?}

\begin{document}

\title{\papertitle}

\author{John Ellis}
\email{john.ellis@cern.ch}
\affiliation{King’s College London, Strand, London, WC2R 2LS, United Kingdom}
\affiliation{Theoretical Physics Department, CERN, Geneva, Switzerland}
  
\author{Malcolm Fairbairn}
\email{malcolm.fairbairn@kcl.ac.uk}
\affiliation{King’s College London, Strand, London, WC2R 2LS, United Kingdom}
   
\author{Juan Urrutia}
\email{juan.urrutia@kbfi.ee}
\affiliation{Keemilise ja Bioloogilise F\"u\"usika Instituut, R\"avala pst. 10, 10143 Tallinn, Estonia}
\affiliation{Departament of Cybernetics, Tallinn University of Technology, Akadeemia tee 21, 12618 Tallinn, Estonia}

\author{Ville Vaskonen}
\email{ville.vaskonen@pd.infn.it}
\affiliation{Keemilise ja Bioloogilise F\"u\"usika Instituut, R\"avala pst. 10, 10143 Tallinn, Estonia}
\affiliation{Dipartimento di Fisica e Astronomia, Universit\`a degli Studi di Padova, Via Marzolo 8, 35131 Padova, Italy}
\affiliation{Istituto Nazionale di Fisica Nucleare, Sezione di Padova, Via Marzolo 8, 35131 Padova, Italy}

\begin{abstract}
We present a new semi-analytical model for the evolution of galaxies and supermassive black holes (SMBHs) that is based on the extended Press-Schechter formalism and phenomenological modelling of star formation. The model yields BH mass-stellar mass relations that reproduce both the JWST and pre-JWST observations. If the efficiency for BH mergers is high the JWST data prefer light seeds while the pre-JWST data prefers heavy seeds. The fit improves for a smaller merger efficiency, $\mathcal{O}(0.1)$, for which both data prefer heavy seeds, while also accommodating the PTA GW background data.
\\~~\\
KCL-PH-TH/2024-55, CERN-TH-2024-184, AION-REPORT/2024-05
\end{abstract}

\maketitle

The origin of supermassive black holes (SMBHs) is (literally) one of the biggest problems in astrophysics and cosmology~\cite{Volonteri:2021sfo}. It has been thrown into the limelight recently by two sets of observations: the discovery by pulsar timing arrays (PTAs) of a (surprisingly strong) stochastic background of nano-Hz gravitational waves (GWs)~\cite{NANOGrav:2023gor,EPTA:2023xxk,Reardon:2023gzh,Xu:2023wog}, commonly thought to have been emitted by SMBH binary systems (though fundamental physics sources are also allowed by the available data~\cite{Ellis:2023oxs}), and the discovery by the JWST~\cite{2023A&A...677A.145U,2023ApJ...953L..29L,2023ApJ...959...39H, Bogdan:2023ilu, 2023Natur.621...51D, Maiolino:2023bpi, 2023arXiv230904614Y,2023ApJ...954L...4K,2024ApJ...964...90S} and other telescopes of a (surprisingly large) population of SMBHs~\cite{2024arXiv240300074L,Pacucci:2023oci} at high redshifts, $z \gtrsim 4$. 

In previous work~\cite{Ellis:2023dgf},  we were able to accommodate the PTA nHz GW data, interpreted as due to SMBH binary systems by using the Extended Press-Schechter (EPS) formalism to model the merger rate of galaxies and assuming a constant probability that each galaxy merger would be accompanied by the merger of their central black holes. Also, we showed in~\cite{Ellis:2024wdh} that JWST and other observations of high-$z$ SMBHs could be described, along with data on low-$z$ inactive galaxies (IGs)~\cite{2015ApJ...813...82R}, by a common relationship between the masses of BHs and the stellar masses of their host galaxies, as well as compatible timescales for the evolution of binaries~\cite{Ellis:2023dgf,Raidal:2024odr}. 

In this paper, we address whether this common stellar mass-BH mass relationship can be understood within the EPS formalism and, if so, the preferred range of BH masses seeding the SMBH assembly process. The co-evolution of SMBHs and galaxies has been studied in the literature with semi-analytical approach using merger trees~\cite{Marulli:2007vm,Dayal:2018gwg,Piana:2020rwb,Trinca:2022txs,2024MNRAS.530.1732P,Toubiana:2024bil,Volonteri:2007ax} and with large-scale N-body and hydrodynamic simulations~\cite{Croton:2005hbr,Schaye:2014tpa,Sijacki:2014yfa,2017MNRAS.470.1121T}. Both of these approaches have a limited mass resolution and are computationally expensive. To overcome these limitations, we compute directly the growth rate of BHs and stellar masses in halo mergers in the EPS formalism. This allows us to track the growth of BHs in small halos and explore the possible importance of light seeds. We also model the accretion of baryonic matter, accounting for supernova (SN) feedback that expels gas from light halos and AGN feedback that heats the gas and prevents star formation. We constrain the star formation rate to accommodate the measured galactic UV luminosity function. Our approach is computationally inexpensive, allowing for systematic scans of the SMBH seeding models.~\footnote{The full evolution from $z=20$ to $z=0$, with $\Delta z=0.1$, takes $\approx 2\, {\rm min}$ on an 8-core M2 Apple Processor, resolving the halos in a range from $M=1\, M_{\odot}$ to $M=10^{20}\, M_{\odot}$ with a logarithmic binning of $\log_{10}[M/M_{\odot}] = 0.25$.}

The combined model of mergers and accretion yields a stellar mass-BH mass relation that is at high masses similar to the empirical relation found in a global fit to the high-$z$, PTA GW and local IG data. Consequently, it is in agreement with the PTA GW observations~\cite{Ellis:2024wdh}, with a preferred merger efficiency  $p_{\rm BH} \sim 0.1$ as also suggested by modeling the PTA background~\cite{Ellis:2023dgf}. The predicted relation depends on the assumed masses of the BH seeds. We find that the JWST data are best described by a scenario with light seeds weighing $\sim 10^2-10^3 M_\odot$ if the BH merger efficiency is high, $p_{\rm BH} \sim 1$, and the rest of SMBH observations can be explained with SMBH seed masses ranging from $10^4-10^5\, M_{\odot}$. If the efficiency of merging is lower, $p_{\rm BH} \sim 0.1$, then heavier seeds with masses $10^4-10^5\, M_\odot$ are preferred by both the JWST and the pre-JWST data. We also find that our model fits are better than power-law fits~\cite{Ellis:2024wdh}. We also discuss the effect of a scatter in the model predictions: the fit quality is improved, but the conclusions are not affected.

\begin{figure*}
    \centering
    \includegraphics[width=\textwidth]{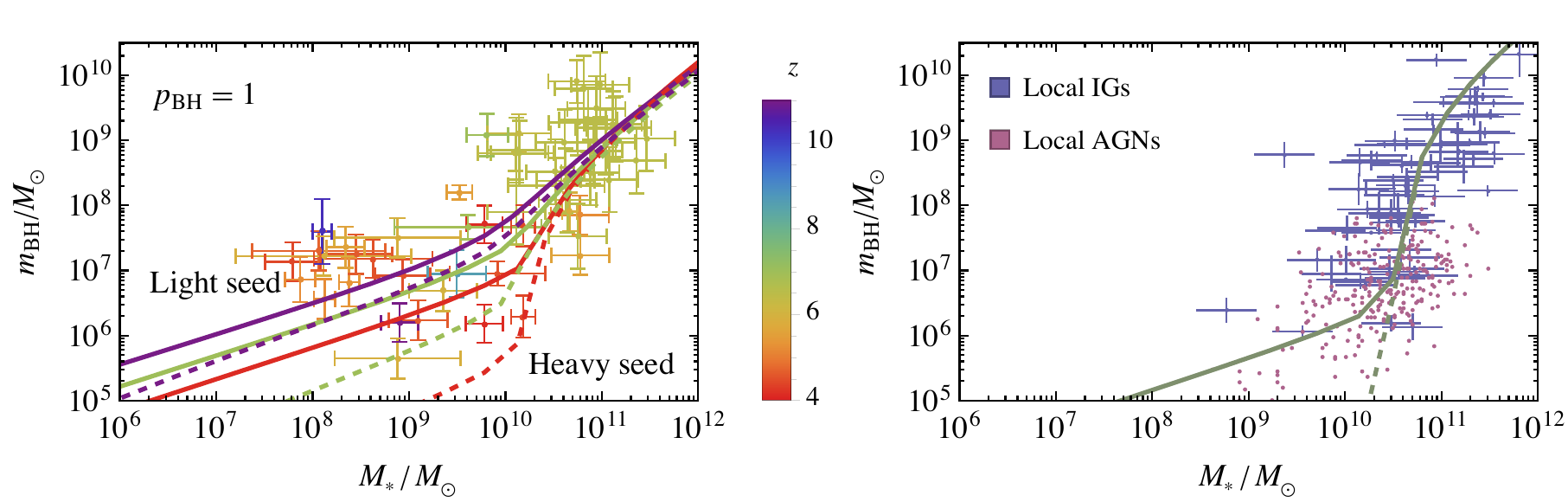}
    \caption{Comparisons of model predictions for the stellar mass-BH mass relation with high-$z$ observations on the left and with low-$z$ observations on the right. The solid lines show the evolution of light seeds ($m_{\rm seed}=100\,M_{\odot}$ with $M_{\rm seed}=3\times10^4\,M_{\odot}$), and dashed lines show the evolution of heavy seeds ($m_{\rm seed}=10^5\, M_{\odot}$ with $M_{\rm seed}=3\times 10^{7}\, M_{\odot}$), both for $p_{\rm BH}=1$.}
    \label{fig:seed_param}
\end{figure*}

\vspace{5pt}\noindent\textbf{Co-evolution of SMBHs and galaxies --} The evolution of the average BH mass and the average stellar mass in the dark matter (DM) halos of mass $M$ is governed by
\bea \label{eq:evolution}
	&\dot M_{\rm BH}(M) = \dot M_{\rm BH}^{\rm merg.}(M) + \dot M_{\rm BH}^{\rm acc.}(M_{\rm BH},M) \,, \\	
	&\dot M_*(M) = \dot M_*^{\rm merg.}(M) + \dot M_*^{\rm sf.}(M) \,,
\eea
where $\dot M_J^{\rm merg.}(M)$: $J = {\rm BH}, *$, denote the growth rates by mergers, $\dot M_{\rm BH}^{\rm acc.}(M_{\rm BH},M)$ is the BH growth rate by accretion, and $\dot M_*^{\rm sf.}(M)$ is the stellar mass growth rate by star formation. The baryonic gas sources the two latter terms, and consequently the evolution of the BH and stellar masses are coupled through the SN and AGN feedback processes that can heat or eject parts of the baryonic gas~\cite{Bower:2005vb,2015MNRAS.452.1502D,2017MNRAS.465...32B,2017MNRAS.468.3935H}. We utilize a phenomenological fit to the star formation rate~\cite{SM} that accounts for the SN and AGN feedbacks and is based on the results of~\cite{2022ApJS..259...20H}, which has been tested successfully with new JWST data~\cite{2023ApJS..265....5H}. Following~\cite{Dayal:2018gwg}, we account consistently for the effect of the SN feedback~\cite{Dayal:2014cda} which strongly suppresses the BH accretion in halos lighter than $M \approx 5\times 10^{11} \,M_\odot$. Moreover, we limit accretion to the Eddington rate. According to Refs.~\cite{Piana:2020rwb,2017MNRAS.465...32B} these are approximately the conditions necessary to explain the UV AGN luminosity function, but rather than an ad-hoc cut on the SMBH mass where accretion is started, it is as a consequence of the interpretation of the star formation fit~\cite{SM}.

We use the EPS formalism to model the growth by mergers. The growth rate by mergers is given by the $\Delta z \equiv z'-z \to 0$ limit of~\cite{SM}
\begin{widetext}
\be \label{eq:mavevol}
    \Delta M_J^{\rm merg.}(M,z,z') = \int_0^M \td M' \bigg|\frac{\td S}{\td M'}\bigg| \frac{M}{M'}  M_J(M',z') \frac{\delta_c(z')-\delta_c(z)}{\sqrt{2\pi [S(M')-S(M)]^3}}e^{-\frac{[\delta_c(z')-\delta_c(z)]^2}{2[S(M')-S(M)]}} - M_J(M,z') \,,
\ee
\end{widetext}
where $J = {\rm BH}, *$, $S(M)$ denotes the variance of the DM perturbations and $\delta_c(z)$ is the threshold DM density contrast for halo formation. We assume the cold DM model and compute $S(M)$ and $\delta_c(z)$ following Refs.~\cite{Dodelson:2003ft,Eisenstein:1997ik}, with the cosmological parameters inferred from the CMB observations~\cite{Planck:2018vyg}. 

We considered also the possibility that not all mergers are efficient at forming a tight BH binary that merges quickly. We parametrize this with an efficiency parameter $0 \leq p_{\rm BH} \leq 1$ that multiplies the BH growth rate by mergers if it is positive. High merger efficiency, $p_{\rm BH}\simeq 1$, might be a good approximation if the SMBH is surrounded by a nuclear star cluster, which is very effective at lowering the orbital timescale~\cite{Mukherjee:2024krx}. However, it might be too optimistic for a more general case and should be considered an effective upper value. Studies of the efficiency of mergers induced by dark matter or three-body effects indicate a reduced BH merger efficiency $p_{\rm BH} \sim \mathcal{O}(0.1)$~\cite{Zhou:2024duc}, although with a big uncertainty.\footnote{This is also consistent, within large uncertainties, with the merger efficiency needed to explain the PTA GW background with inspiralling SMBH binaries~\cite{Ellis:2023dgf}.}

\begin{figure}
  \centering
    \includegraphics[width=0.49\columnwidth]{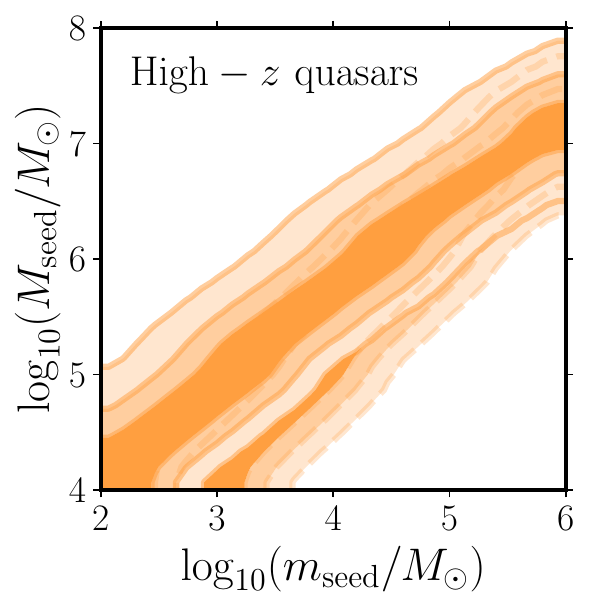} 
    \includegraphics[width=0.49\columnwidth]{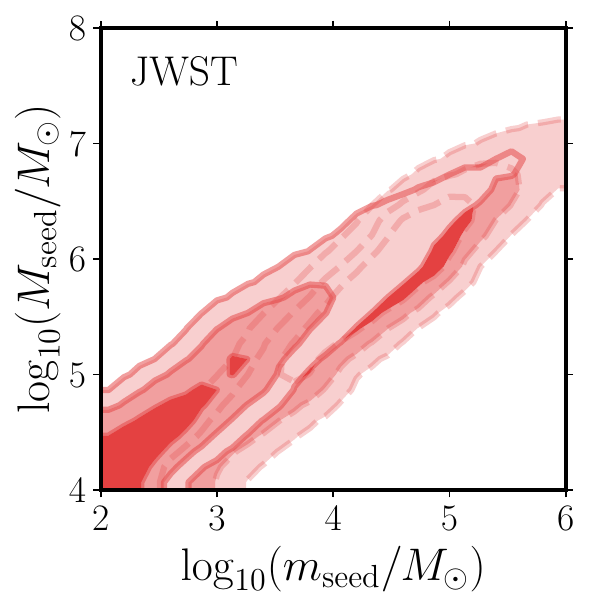} \\ 
    \includegraphics[width=0.49\columnwidth]{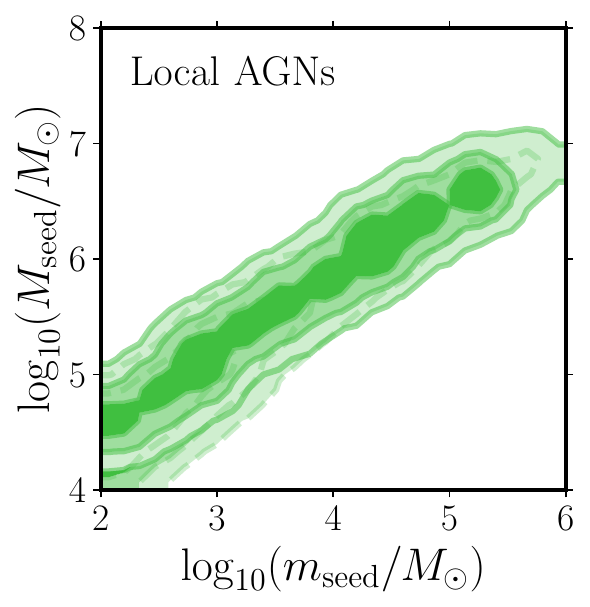} 
    \includegraphics[width=0.49\columnwidth]{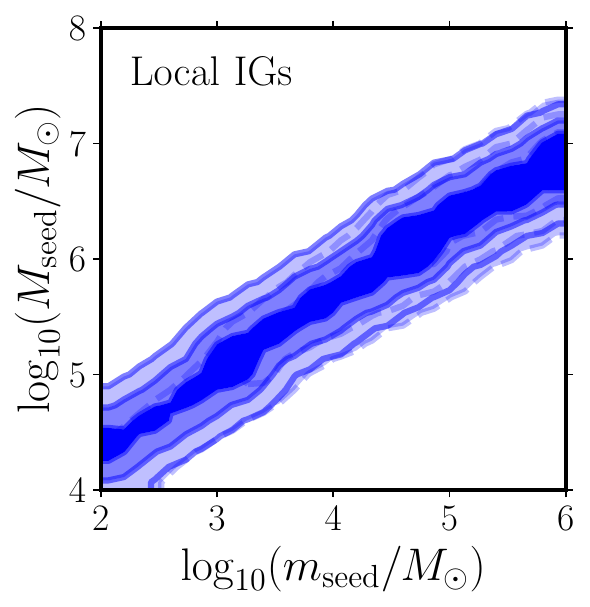}
    \caption{Fits to stellar mass-BH mass data as functions of the seed BH mass $m_{\rm seed}$ and minimal halo mass $M_{\rm seed}$, where the seeds are inserted at $z_{\rm seed} = 20$ with $p_{\rm BH}=1$ in solid and $p_{\rm BH}=0.1$ in dashed. {\it Upper left:} High-$z$ quasar data compiled in~\cite{2021ApJ...914...36I}. {\it Upper right:} High-$z$ observations with JWST~\cite{2023A&A...677A.145U,2023ApJ...953L..29L,2023ApJ...959...39H, Bogdan:2023ilu, 2023Natur.621...51D, Maiolino:2023bpi, 2023arXiv230904614Y}. {\it Lower left:} Low-z AGNs as compiled in~\cite{2015ApJ...813...82R}. {\it Lower right:} Low-$z$ IGs~\cite{2015ApJ...813...82R}.}
    \label{fig:fits_to_SMBH_data}
\end{figure}

We initiate the evolutionary process by planting a seed of mass $m_{\rm seed}$ at some redshift $z_{\rm seed}$ in every halo that is heavier than some minimal mass $M_{\rm seed}$, and evolve the BH masses and stellar masses by solving numerically the coupled equations~\eqref{eq:evolution}. Note that we evolve only the expected BH mass, $M_{\rm BH} = \int \td P(m_{\rm BH}) \,m_{\rm BH}$, and do not include the inevitable scatter in the present computations. We therefore estimate the occupation fractions with Dirac delta functions
\be \label{eq:Pocc}
    \frac{\td P(m_{\rm BH}|M_*,z)}{\td m_{\rm BH}} = \frac{M_{\rm BH}(M_*,z)}{m_{\rm BH}} \delta\left[m_{\rm BH} - m_{\rm BH}(M_*,z)\right] \,,
\ee
where
\be
    m_{\rm BH}(M_*,z) = \max\left[m_{\rm seed},M_{\rm BH}(M_*,z)\right]
\ee
corresponds to the proper BH mass. In order to explore how a spread the results could affect the comparison with the data, we can replace the Dirac delta function in~\eqref{eq:Pocc} with a lognormal distribution, $\mathcal{N}(\log_{10} m_{\rm BH} |\log_{10} m_{\rm BH}(M_*,z),\sigma)/m_{\rm BH}$, where $\mathcal{N}(x|\bar{x},\sigma)$ denotes the Gaussian distribution with mean $\bar x$ and the variances $\sigma^2$, as discussed below.

\vspace{5pt}\noindent\textbf{Implications of the data --} We use the stellar mass-BH mass data from~\cite{2023A&A...677A.145U,2023ApJ...953L..29L,2023ApJ...959...39H, Bogdan:2023ilu, 2023Natur.621...51D, Maiolino:2023bpi, 2023arXiv230904614Y,2023ApJ...954L...4K,2024ApJ...964...90S} for the JWST AGNs, the high-redshift quasar masses compiled in~\cite{2021ApJ...914...36I} and the data on local AGNs and IGs (where SMBH masses are measured dynamically) compiled in~\cite{2015ApJ...813...82R}. Examples of our model predictions for the stellar mass-BH mass correlation for $p_{\rm BH} = 1$ are shown alongside the data in Fig.~\ref{fig:seed_param}. The benchmark cases shown correspond to examples with light seeds ($m_{\rm seed}=100\,M_{\odot}$ and $M_{\rm seed}=3\times10^4\,M_{\odot}$) and heavy seeds ($m_{\rm seed}=10^5\, M_{\odot}$ with $M_{\rm seed}=3\times 10^{7}\, M_{\odot}$). The bend in the model predictions at $M_* \sim 10^{10} M_\odot$ arises because SN feedback expels all of the gas remaining after star formation below a critical halo mass, preventing efficient BH accretion. Therefore growth via mergers is the dominant growth channel for BHs in such light galaxies. Light-seed scenarios have enough light BHs to fuel the growth of the AGNs observed by JWST, but the heavy-seed model predictions undershoot the JWST observations for $M_* \lesssim 10^{9} M_\odot$. The high-$z$ quasar data with $M_* \gtrsim 10^{10} M_\odot$ compiled in~\cite{2021ApJ...914...36I} are, instead, fitted well by either light or heavy seeds. In the right panel of Fig.~\ref{fig:seed_param} we see that both the low-$z$ IGs and the low-$z$ AGN data are fitted well thanks to the steep fall of the model predictions at $M_* \sim 10^{11} - 10^{10} M_\odot$ arising from the SN feedback.

\begin{figure*}
    \centering
    \includegraphics[width=\textwidth]{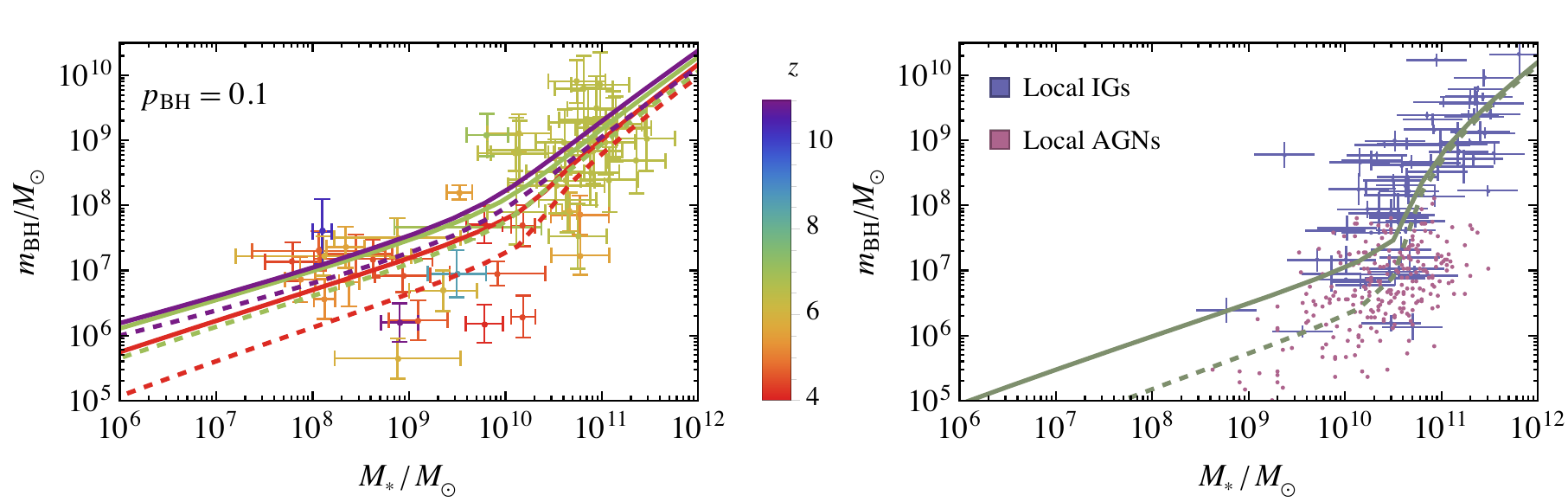}
    \caption{Comparisons of model predictions for the stellar mass-BH mass relation with high-$z$ observations on the left and with low-$z$ observations on the right. The solid lines  show the best fit for $p_{\rm BH} = 0.1$ to the JWST data, and the dashed lines show the best fit to the rest of the SMBH data. The best-fit parameter values are given in Table~\ref{tab:comparison}.}
    \label{fig:best_fit}
\end{figure*}

\begin{figure}
  \centering
    \includegraphics[width=0.49\columnwidth]{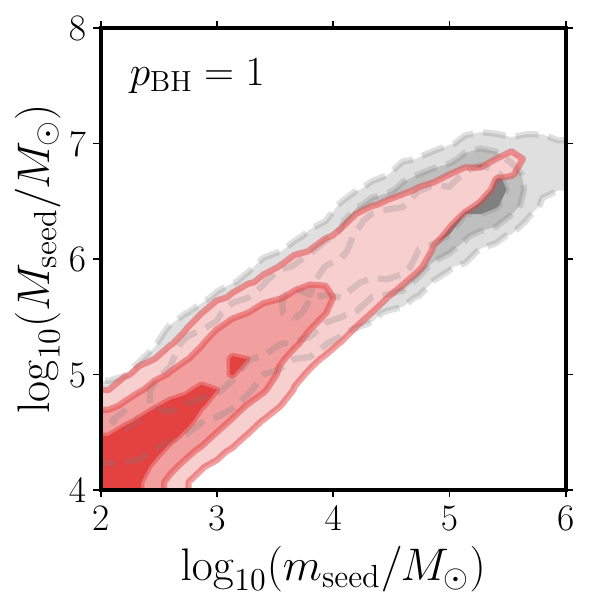}
    \includegraphics[width=0.49\columnwidth]{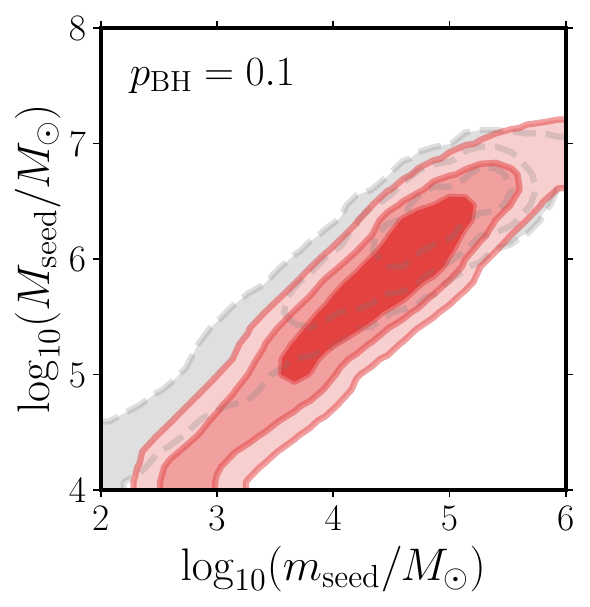}
    \vspace{-2mm}
    \caption{Fits to stellar mass-BH mass data: the red posteriors show the fit to the JWST data and the gray posteriors show the fit combining the low-$z$ data and the high-$z$ quasar data.}
    \label{fig:global_SMBH_data}
\end{figure}
 
We compare the qualities of our model predictions to the stellar mass-BH mass observations using the likelihood function
\begin{widetext}
\be
    \mathcal{L}(\vect{\theta}) \propto \prod_j \! \int \!\td \log_{10} \!m_{\rm BH}  \td \log_{10} M_*  \frac{\td P(m_{\rm BH}|M_*,z_j,\vect{\theta})}{\td \log_{10} m_{\rm BH}} \mathcal{N}(\log_{10} m_{\rm BH} | \log_{10} m_{{\rm BH},j}, \sigma_{{\rm BH},j}) \mathcal{N}(\log_{10} M_*|\log_{10} M_{*,j},\sigma_{*,j}) ,
\ee
\end{widetext}
where the product is over the data points and the Gaussian probability densities $\mathcal{N}(x|\bar{x},\sigma)$ account for the uncertainties in the measurements of the BH masses and the galaxy stellar masses. Fixing all the parameters associated with star formation by observations~\cite{SM}, the free parameters are those associated with the formation of the first generation of massive BHs and the efficiency factor of the BH growth by mergers: 
\be
    \vect{\theta} = \left( m_{\rm seed}\, , M_{\rm seed}\, , z_{\rm seed}, p_{\rm BH} \right) \,.
\ee 
We find that varying $z_{\rm seed} \in(15-20)$ does not significantly affect the likelihood. The following results have been calculated for fixed $z_{\rm seed} = 20$.

We have performed scans over $m_{\rm seed}$ and $M_{\rm seed}$ for fixed values of $p_{\rm BH}=1$ and $p_{\rm BH}=0.1$, as motivated above. In Fig.~\ref{fig:fits_to_SMBH_data} we show the posterior distributions separately for different data sets ignoring a possible spread in the mass relation. The upper left plot displays results for fits to the high-$z$ quasar data compiled in~\cite{2021ApJ...914...36I}. We see that, whilst these data favour a strong correlation between $m_{\rm seed}$ and $M_{\rm seed}$, they do not favour any particular range of seed masses. This reflects the observation that the predictions of both the light- and heavy-seed models pass through the high-$z$ SMBH data compiled in~\cite{2021ApJ...914...36I}. On the other hand, as seen in the upper right plot, the JWST data show a preference for light seeds with masses $m_{\rm seed} \lesssim 10^3 M_\odot$ if $p_{\rm BH} = 1$ and heavier seeds with masses $10^4 M_\odot \lesssim m_{\rm seed} \lesssim 10^5 M_\odot$ if $p_{\rm BH} = 0.1$. The low-$z$ observations, shown in the lower panels, also show strong correlations between the seed halo mass and the seed BH mass, with the contours scaling roughly as $M_{\rm seed} \propto m_{\rm seed}^{0.6}$, but do not by themselves constrain significantly the seed BH mass. 

In Fig.~\ref{fig:global_SMBH_data} we compare the fit combining the low-$z$ data and the high-$z$ quasar data with the fit to the JWST high-$z$ data. We note that the trend of the fit to the high-$z$ quasar data is slightly steeper than for the low-$z$ data, $M_{\rm seed} \propto m_{\rm seed}^{0.8 - 1}$, which is why the fit combining the low-$z$ data and the high-$z$ quasar data shown in gray in Fig.~\ref{fig:global_SMBH_data} has a preferred seed mass range. The JWST fit shows a preference for lighter seeds than the combined fit. This preference is stronger for $p_{\rm BH} = 1$ than for $p_{\rm BH} = 0.1$, but even for $p_{\rm BH} = 1$ the fits overlap significantly within the $2\sigma$ confidence level. We show in Fig.~\ref{fig:best_fit} the evolution of the $(m_{\rm seed}, M_{\rm seed})$ relation for the best fits to JWST data and the rest of the SMBH data for $p_{\rm BH}=0.1$: parameter values are given in Table~\ref{tab:comparison}. We see how the JWST-only fit, shown as solid lines, traverses the data and has the correct $z$ dependence. On the other hand, the fit to the rest of the data, shown as dashed lines, reproduces better the local data. 

The JWST and other datasets exhibit significant scatter, with objects both above and below the model predictions, while the fits presented in Figs.~\ref{fig:global_SMBH_data} and \ref{fig:fits_to_SMBH_data} assume zero spread in the model predictions. To explore whether our results are sensitive to spread in our model predictions, we have repeated the analysis including a lognormal spread with $\sigma = 0.5$ in the predicted SMBH mass. The fits are shown in the Supplemental Material~\cite{SM}. Our overall conclusions do not change, but the preference of the fit moves to slightly smaller seed masses and the quality of the fit improves. We have collected best fits from the different scans in Table~\ref{tab:comparison}, which also shows the best fits obtained for a power-law fit as in~\cite{Ellis:2024wdh}. Impressively, our model provides significantly better fits than the power-law fit with the same number of degrees of freedom. Moreover, we see that a spread in the model predictions clearly improves the quality of the fit and the data prefer a lower merger efficiency of $p_{\rm BH} = 0.1$ over $p_{\rm BH} = 1$, independently of the spread. However, we also see that, in particular for $p_{\rm BH} = 0.1$, the seed BHs need to be relatively heavy compared to the minimal halo masses where they are planted, on the other hand, the JWST data become more compatible with the rest of the SMBH data.

\begin{table*}
    \centering
    \resizebox{0.85\textwidth}{!}{ 
        \begin{tabular}{c|c|c|c|c|c|c}
             & $p_{\rm BH}=1$  & $p_{\rm BH}=0.1$ & $p_{\rm BH}=1$ & $p_{\rm BH}=0.1$ & power-law & power-law \\
             & no spread & no spread & $\sigma=0.5$ & $\sigma=0.5$ & no spread & $\sigma=0.5$ \\
            \hline
            JWST & $m_{\rm seed}=10^2\,M_{\odot}$ & $3\times 10^{4}\,M_{\odot}$ & $10^2\,M_{\odot}$ & $2\times10^4\,M_{\odot}$ & $a=8.5$ & $8.5$\\
            & $M_{\rm seed}=10^4\,M_{\odot}$ & $5\times 10^5\,M_{\odot}$ & $10^4\,M_{\odot}$ & $3\times10^5\,M_{\odot}$ & $b=0.64$ & $0.64$\\
            & $-2\log{\mathcal{L}}=200$ & $200$ & $70$ & $60$ & $-2\log{\mathcal{L}}=230$ & $140$ \\
            \hline
            pre-JWST & $3\times10^4\,M_{\odot}$ & $6\times10^4\,M_{\odot}$ & $5\times10^3\,M_{\odot}$ & $4\times10^3\,M_{\odot}$ & $a=8.3$ & $8.0$\\ & $10^6\,M_{\odot}$ & $2\times10^6\,M_{\odot}$ & $3\times10^{5}\,M_{\odot}$ & $2\times10^5\,M_{\odot}$ & $b=1.8$ & $1.4$\\
            & $1080$ & $990$ & $780$ & $650$ & $1100$ & $970$\\
            \hline
            all data& $3\times10^4\,M_{\odot}$ & $2\times10^5\,M_{\odot}$ & $5\times10^3\,M_{\odot}$ & $10^5\,M_{\odot}$ & $a=8.3$ & $8.0$\\
            &$10^6\,M_{\odot}$ & $3\times10^6\,M_{\odot}$ & $3\times10^5\,M_{\odot}$ & $2\times 10^6\,M_{\odot}$ & $b=1.4$ & $1.0$\\
            & $1290$ & $1190$ & $850$ & $720$ & $1820$ & $1290$\\
        \end{tabular}
    }
    \caption{{\it Leftmost 4 columns}: Compilation of the best fit and log-likelihood values of the model parameters described in the text.
    {\it Rightmost 2 columns}: The best fits for a power-law mean BH mass-stellar mass relation $m_{\rm BH} = 10^a M_\odot (M_*/10^{11}M_\odot)^b$ with a lognormal spread defined as in~\cite{Ellis:2024wdh} are shown for comparison.}
    \label{tab:comparison}
\end{table*}

Ref.~\cite{Lambrides:2024ugh} (see also~\cite{Pacucci:2024tws}) raised the possibility that the masses of at least some of the JWST SMBHs~\cite{Maiolino:2023bpi} may have been overestimated. It has been argued that the non-detection of measurable X-rays or UV emission lines suggests that these SMBHs may be lower-mass objects accreting at super-Eddington rates. As seen in Fig.~2 of~\cite{Lambrides:2024ugh}, this argument would reduce the estimated masses of some SMBHs observed with JWST~\cite{Maiolino:2023bpi} by perhaps an order of magnitude. This would shift the preference of the JWST data towards seeds planted in heavier halos.

\vspace{5pt}\noindent\textbf{Conclusions --} We have presented in this paper a new semi-analytical model of the co-evolution of SMBHs and their host galaxies. This model is based on estimates of the growth rate of BH masses and the galaxy stellar masses derived from the EPS formalism and on phenomenological modeling of star formation, SN and AGN feedback processes and BH accretion. In particular, we include suppression of BH accretion in halos lighter than $\approx 5\times 10^{11}\,M_\odot$ due to SN feedback, and a parameter that describes how efficient the halo mergers are at forming a tight BH binary. Our approach involves solving numerically coupled differential equations for the evolution of the SMBH masses and the stellar masses, and is much faster than approaches based on merger trees or numerical simulations. This has allowed us to perform a systematic study of different SMBH seeding scenarios including scans of the model parameter space.

Our model predicts BH mass-stellar mass relations that are compatible with the observations. We have found that the heavy seed scenarios where BHs of mass $\sim 10^5 M_\odot$ are planted in halos heavier than $\sim 10^7 M_\odot$ undershoot the JWST observations because the BHs in halos lighter than $\approx 5\times 10^{11}\,M_\odot$ grow dominantly by mergers but there are not enough BHs to merge with. Scenarios with lighter seeds planted into lighter halos are preferred by JWST as they allow for more efficient growth of the BHs by mergers.

We have performed a likelihood analysis scanning over the seed BH masses and the masses of the smallest halos into which they are planted. As seen in Table I, our analysis suggests that the JWST data prefer lighter seeds $m_{\rm seed}\lesssim 10^3 M_\odot$ if the merging efficiency is high, $p_{\rm BH} = 1$, but the preference moves to higher masses $10^4 M_\odot \lesssim m_{\rm seed}\lesssim 10^5 M_\odot$ for a lower merger efficiency of $p_{\rm BH} = 0.1$. The other data from local observations and observations of high-$z$ quasars instead prefer the higher seed mass range, $10^4 M_\odot \lesssim m_{\rm seed}\lesssim 10^5 M_\odot$, for both $p_{\rm BH} = 1$ and $p_{\rm BH} = 0.1$. The fits have two non-trivial successes. First, the fits are better for $p_{\rm BH}=0.1$, which aligns with a fit to the PTA GW background~\cite{Ellis:2023dgf}. Secondly, our model provides a much better fit than those obtained with a simple power-law parametrization~\cite{Ellis:2024wdh}, as also seen in Table~I.

Our model does not resolve the spread in the BH masses but we have explored the effects of the spread by parametrizing it. We have found that including the spread significantly improves the fit to the observations, as also seen in Table~I. This is expected given that the data exhibit significant scatter. The inclusion of the spread in the model predictions also prefers slightly lower seed masses but does not change the overall conclusions.

Building on the success of our model framework to study the scatter in the range of SMBH masses for any given value of the stellar mass is clearly a priority. Another interesting extension of this paper would be to calculate the magnitude of the stochastic GW background due to BH binaries to the higher frequencies than those explored by the PTAs that could be explored by LISA~\cite{2017arXiv170200786A} and deciHz experiments~\cite{Badurina:2019hst,AEDGE:2019nxb,Kawamura:2020pcg}. However, these explorations lie beyond the scope of this paper.

\newpage

\begin{acknowledgments}
\vspace{5pt}\noindent\emph{Acknowledgments --} This work was supported by the Estonian Research Council grants PRG803, PSG869, RVTT3 and RVTT7 and the Center of Excellence program TK202. The work of J.E. and M.F. was supported by the United Kingdom STFC
Grants ST/T000759/1 and ST/T00679X/1. The work of V.V. was partially supported by the European Union's Horizon Europe research and innovation program under the Marie Sk\l{}odowska-Curie grant agreement No. 101065736.
\end{acknowledgments}

\bibliography{refs}

\newpage
\clearpage
\onecolumngrid

\begin{center}
\textbf{\large \papertitle} \\ 
\vspace{0.06in}
{John Ellis, Malcolm Fairbairn, Juan Urrutia and Ville Vaskonen} \\ 
\vspace{0.1in}
{SUPPLEMENTAL MATERIAL}
\vspace{0.1in}
\end{center}

\setcounter{equation}{0}
\setcounter{figure}{0}
\setcounter{section}{0}
\setcounter{table}{0}
\setcounter{page}{1}
\makeatletter
\renewcommand{\theequation}{S\arabic{equation}}
\renewcommand{\thefigure}{S\arabic{figure}}
\renewcommand{\thetable}{S\arabic{table}}

\section{Going beyond halo merger trees} 
\label{sec:BHtrees}

Although the EPS formalism is a convenient way to derive halo merger trees and thereby BH histories it has its drawbacks, mainly that halos with masses below $\sim10^8\,M_{\odot}$ are too abundant and computationally expensive to include (although this can be improved by an adaptive resolution~\cite{Toubiana:2024bil}). In this paper, we aim to overcome these drawbacks by applying the EPS formalism directly to the evolution of SMBHs without invoking the halo merger tree step. The key insight behind this approach is that \textit{a SMBH inside a halo on average has had the same history as the average mass element of the halo.} This implies that the statistics from the random paths for the mass elements of a halo at some $(M,z)$ are the same as for the SMBH inside that halo, which is the key assumption used when their evolution is calculated using halo merger trees. 

In the cold dark matter model, the variance of matter perturbations $S(M)$ is a monotonically decreasing function of the halo mass $M$: $S(M) \to 0$ as $M\to \infty$. Thus, we can replace $M$ by $S \equiv S(M)$ and the matter overdensity field $\delta$ around each spatial point evolves along a path in the $(S(M),\delta)$ plane that originates from $(0,0)$. The first crossing of the path across a threshold value $\delta_c(z)$ determines the halo mass $M$ around that point at redshift $z$. {We compute the threshold $\delta_c(z)$ as in~\cite{Dodelson:2003ft} and the variance $S(M)$ following~\cite{Eisenstein:1997ik}, with the cosmological parameters fixed by the CMB observations~\cite{Planck:2018vyg}.}

Assuming that the fluctuations are Gaussian, the probability a path to get to the point $(S,\delta)$ is $P(\delta|S) = e^{-\delta^2/2S}/\sqrt{2\pi S}$ and the probability that the first crossing is at variance larger than $S$ is given by 
\bea \label{eq:FC}
    P_{\rm FC}(S,z) &= 1 - \int_{-\infty}^{\delta_c(z)}\td \delta\, \left[P(\delta|S) - P(2\delta_c(z)-\delta|S) \right] \\
    &= 1 - {\rm erf}\left[\frac{\delta_c(z)}{\sqrt{2 S}}\right] \,. 
\eea
This is the cumulative mass fraction in halos lighter than $M$ and the EPS halo mass function (HMF) is simply
\bea \label{eq:EPSHMF}
    \frac{\td n(z)}{\td M} =& \, \frac{\rho_{\rm DM}}{M} \left|\frac{\td S}{\td M}\right| \frac{\td P_{\rm FC}(z)}{\td S} \\ 
    =& \, \frac{\rho_{\rm DM}}{M} \left|\frac{\td S}{\td M}\right| \frac{\delta_c(z)}{\sqrt{2\pi S^3}} e^{-\frac{\delta_c(z)^2}{2S}} \,,
\eea
where $\rho_{\rm DM}$ is the present average DM energy density.

\begin{figure}
    \centering
    \includegraphics[width=0.5\textwidth]{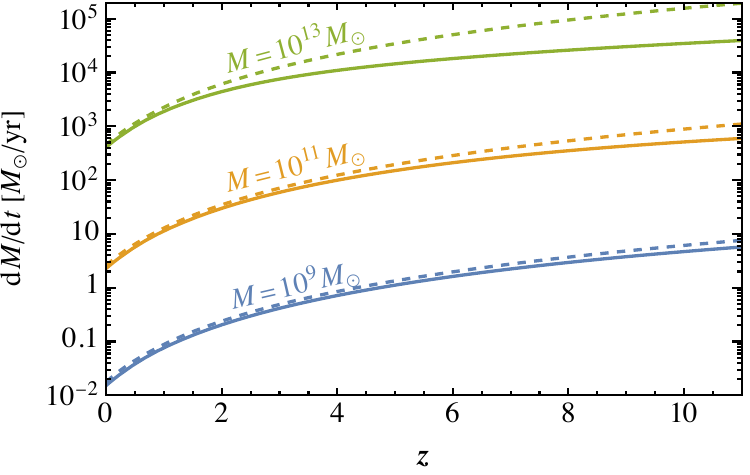}
    \vspace{-3mm}
    \caption{Calculations of the growth rates of DM halos. The solid curves are obtained using Eq.~\eqref{eq:Mdot} and used in our analysis. The dashed curves show for comparison the estimates derived in~\cite{Correa:2014xma}, which are based on finding the average mass of the main progenitor.}
    \label{fig:Mdot}
\end{figure}

If the path is not initiated in $(0,0)$ but instead at $(S,\delta)$, the probability to get to the point $(S',\delta')$, $S' > S$ becomes $P(\delta' | \delta, S, S') = P(\delta' - \delta | S' - S)$. The probability for a path starting at $(S,\delta)$ to cross the threshold $\delta' = \delta_c(z') > \delta = \delta_c(z)$ at variance larger than $S'$ is, in the same way as in~\eqref{eq:FC}, given by
\bea \label{eq:FC_2}
    P_{\rm FC}(\delta',S'|\delta,S) = 1 - {\rm erf}\left[\frac{\delta'-\delta}{\sqrt{2(S'-S)}}\right] \,,
\eea
and the probability for the path to cross the threshold $\delta'$ in the range $(S',S'+\td S')$ is
\bea \label{eq:FC_2}
    p_{\rm FC}(\delta',S'|\delta,S) \td S' &= \frac{\td P_{\rm FC}(\delta',S'|\delta,S)}{\td S'} \td S' \\
    &= \frac{\delta'-\delta}{\sqrt{2\pi(S'-S)^3}}e^{-\frac{(\delta'-\delta)^2}{2(S'-S)}} \,\td S'\, .
\eea
From this one can compute what fraction of the mass $M$ of a halo at $z$ on average comes from a halo of mass $M'$ in the past, $z' > z$. If instead, we want to compute how the mass of the halo $M'$ at $z'$ is going to be distributed in heavier halos $M>M'$ in some later time $z < z'$, we need to invert the conditional probability~\eqref{eq:FC_2}. This gives the probability that a path that starts at $(\delta',S')$ crosses the threshold $\delta$ in the range $(S,S+\td S)$ (see Eq.~(2.16) in~\cite{Lacey:1993iv}):
\bea
    p_{\rm FC}(\delta,S|\delta',S') \td S =& \,p_{\rm FC}(\delta',S'|\delta,S)\frac{p_{\rm FC}(\delta,S|0,0)}{p_{\rm FC}(\delta',S'|0,0)} \td S \\
    =& \,p_{\rm FC}(\delta',S'|\delta,S) \left(\frac{S'}{S}\right)^{\frac32} \left(\frac{\delta}{\delta'}\right) e^{\frac{\delta'^2}{2S'}-\frac{\delta^2}{2S}} \, \td S \,.
\eea
A change of variables then gives the probability that the halo whose mass at $z'$ is $M'$ ends up being a part of a halo in the mass range $(M,M+\td M)$ at $z<z'$:
\be
    \td P(M,z|M',z') \equiv p_{\rm FC}(\delta,S|\delta',S') \bigg|\frac{\td S}{\td M}\bigg| \td M \,.
\ee 
This allows us to compute the evolution of the HMF forward in time from $z' > z$ to $z$ as 
\be
    \frac{\td n(z)}{\td M} = \int_0^M\td M' \frac{\td n(z')}{\td M'} \frac{M'}{M} \frac{\td P(M,z|M',z')}{\td M} \,.
\ee
It is easy to check that this formula holds by inserting the HMF~\eqref{eq:EPSHMF} in the right-hand side and performing the integral. We estimate the growth rate of the DM halos by computing the expected growth of the halo through mergers with smaller halos:
\be \label{eq:Mdot}
	\dot M = (1+z) H(z)  \lim_{\Delta z \to 0} \frac{1}{\Delta z} \left[\frac{\int_{M}^{2M} \! \td \tilde M\, \tilde M \frac{\td P(\tilde M,z|M,z+\Delta z)}{\td \tilde M}}{\int_{M}^{2M} \! \td \tilde M\, \frac{\td P(\tilde M,z|M,z+\Delta z)}{\td \tilde M}} - M \right] \,.
\ee
We show in Fig.~\ref{fig:Mdot} DM halo growth rates calculated  using Eq.~\eqref{eq:Mdot} (solid lines) and, for comparison, estimates derived in~\cite{Correa:2014xma}, which is based on finding the average mass of the main progenitor. The agreement is generally good over the range of halo masses $M$ relevant for our study.

\section{SMBHs and baryonic physics} 
\label{sec:baryons}

The average BH mass and the stellar mass evolve according to the coupled equations
\bea
	&\dot M_{\rm BH}(M, p_{\rm BH}) = \dot M_{\rm BH}^{\rm merg.}(M, p_{\rm BH}) + \dot M_{\rm BH}^{\rm acc.}(M_{\rm BH},M) 
    \,, \\	
	&\dot M_*(M) = \dot M_*^{\rm merg.}(M) + \dot M_*^{\rm sf.}(M) \,,
\eea
where $\dot M_J^{\rm merg.}(M)$, $J =\left( {\rm BH},\,*\,, {\rm gas}\right)$, denotes the growth rate by mergers, $\dot M_{\rm BH}^{\rm acc.}$ denotes BH accretion rate, and $\dot M_*^{\rm sf.}$ is the star formation rate. The baryonic matter (stars and gas) and the SMBH have the same progenitors as the DM halo that hosts them. So, considering only mergers, their expected masses, $M_J$, evolve from $z'$ to $z$ in a halo of mass $M$ as 
\bea \label{eq:mavevol}
    M_J(M,z') + \Delta M_J^{\rm merg.}(M,z,z') =& \left[\frac{\td n(z)}{\td M}\right]^{-1}\!\! \int_0^M \td M'\,\frac{\td n(z')}{\td M'} \frac{\td P(M,z|M',z')}{\td M} M_J(M',z') \\
    =& \int_0^M \td M' \bigg|\frac{\td S}{\td M'}\bigg| \frac{M}{M'} M_J(M',z') \frac{\delta_c(z')-\delta_c(z)}{\sqrt{2\pi [S(M')-S(M)]^3}}e^{-\frac{[\delta_c(z')-\delta_c(z)]^2}{2[S(M')-S(M)]}} \,.
\eea
The growth rate by mergers is given by the limit
\be \label{eq:Jmerg}
	\dot M_J^{\rm merg.}(M) = (1+z) H(z)  \lim_{\Delta z \to 0} \frac{\Delta M_J^{\rm merg.}(M,z,z+\Delta z)}{\Delta z} \,.
\ee

For the growth rate of BHs by mergers, $\dot M_{\rm BH}^{\rm merg.}(M, p_{\rm BH})$, we considered the possibility that not all mergers are efficient at forming a tight binary and the consequent BH is left orbiting at very high separations and it is no longer efficient at accreting or merging. In that case, we considered 
\be
\dot M_{\rm BH}^{\rm merg.}(M, p_{\rm BH}) = 
\begin{cases}
    M_{\rm BH}^{\rm merg.}(M) & <0 \\
    p_{\rm BH} \,M_{\rm BH}^{\rm merg.}(M) & \geq 0
\end{cases} \,.
\ee
The part $1-p_{\rm BH}$ of the accreted SMBH mass can be thought of as an SMBH that is left wandering in the galaxy or becomes ejected and does not play any significant role in future mergers. 

\begin{figure}
    \centering
    \includegraphics[height=0.32\textwidth]{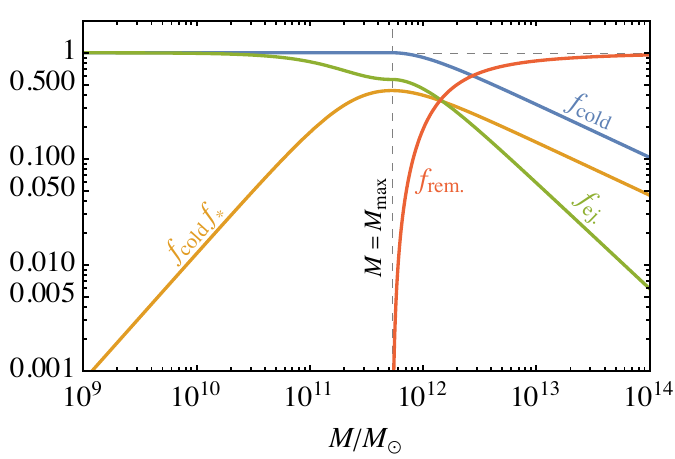} \hspace{2mm}
    \includegraphics[height=0.32\textwidth]{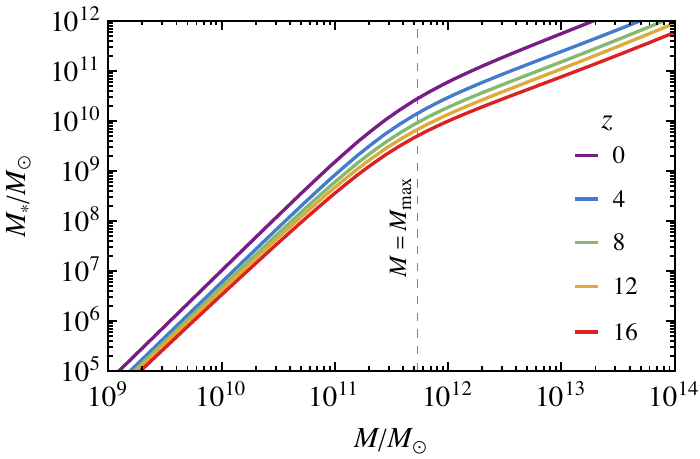}
    \vspace{-2mm}
    \caption{{\it Left panel}: Baryon fractions as functions of the DM halo mass $M$, where $f_{\rm ej.}$ is the cold gas fraction that is ejected from the halo by SNe, $f_{\rm cold}$ is the fraction of the remaining gas that is cold, $f_{\rm rem}(M)$ is the remaining fraction of gas that is remaining after star formation and SN feedback and $f_*$ is the fraction of cold gas that is used for star formation. {\it Right panel}: The evolution of the halo mass-stellar mass relation given by the star formation rate~\eqref{eq:SFR}, the EPS estimate for the halo growth~\eqref{eq:Mdot} and the growth of stellar masses by mergers~\eqref{eq:Jmerg}.}
    \label{fig:fractions}
\end{figure}

Our aim in the following is to utilize phenomenological fits as much as possible while combining the different components in a physically consistent manner. We start from the star formation rate found in \cite{2022ApJS..259...20H}, based on fitting the UV luminosity function
\bea \label{eq:SFR}
    \dot M_*^{\rm sf.}(M,z) &= \frac{6.4\times 10^{-2}[0.53\tanh{\left(0.54(2.9-z)\right)}+1.53]}{(M/M_{\rm crit})^{-1.2}+(M/M_{\rm crit})^{0.5}}  \dot M  \\
    &= f_*(M,z) f_{\rm cold}(M) f_B \dot M
\eea
with $M_{\rm{crit}}=10^{11.5}\, M_{\odot}$. Stars are formed from the cold baryonic gas. In the second line of~\eqref{eq:SFR} we have factorized the fit into the fraction of cold gas,
\be
    f_{\rm cold}(M) \equiv
    \begin{cases}
        1 \,, & M < M_{\rm max}\\
        \frac{1.83}{(M/M_{\rm crit})^{-1.2}+(M/M_{\rm crit})^{0.5}} \,, & M \geq M_{\rm max}
    \end{cases} \,,
\ee 
and the effective star formation efficiency, which determines the fraction of cold gas that is converted to stars,
\be
    f_*(M,z) \equiv \left[ 0.33 + 0.12\tanh{\left(0.54(2.9-z)\right)} \right] \times
    \begin{cases}
        \frac{1.83}{(M/M_{\rm crit})^{-1.2}+(M/M_{\rm crit})^{0.5}} \,, & M<M_{\rm max} \\
        1 \,, & M\geq M_{\rm max}
    \end{cases} \,.
\ee 
The mass $M_{\rm max} = 5.3\times 10^{11} M_\odot$ corresponds to the maximum of $\dot M_*^{\rm sf.}(M,z)$. Finally, we define the fraction of the cold gas that is ejected from the halo by supernovae (SNe) as
\be
f_{\rm ej.}(M,z) \equiv \big( 1-f_*(M,z) \big) \times
\begin{cases}
    1 \,, & M<M_{\rm max} \\
    f_{\rm cold}(M,z)^2 \,, & M\geq M_{\rm max}
\end{cases} \,.
\ee  
This means that the cold gas used for star formation is sufficient for the feedback from the resulting SNe to expel the remaining gas. Star formation reaches its maximum at $M = M_{\rm max}$, and the SNe are no longer able to eject all the gas at $M > M_{\rm{max}}$. Therefore, for $M > M_{\rm{max}}$ there is still gas available to fuel AGN activity, which has the effect of heating the gas in the galaxy, leading to a downturn in the star formation, as illustrated in the left panel of Fig.~\ref{fig:fractions}. The evolution of stellar masses is fixed by the star formation rate~\eqref{eq:SFR}, the EPS estimate for the halo growth~\eqref{eq:Mdot} and the growth of stellar masses by mergers~\eqref{eq:Jmerg}. We show the resulting evolution of the halo mass-stellar mass relation in the right panel of Fig.~\ref{fig:fractions}.

The amount of gas left in the halo after the star formation period is available for accretion by the SMBH. Following~\cite{Dayal:2018gwg}, we estimate the BH accretion rate as
\be
    \dot M_{\rm BH}^{\rm acc.}(M_{\rm BH},M) = \min\left\{f_{\rm rem.}(M,z) f_B (f_1^{\rm acc.}\dot M + f_2^{\rm acc.} M), \, f_{\rm Edd.}\dot{M}_{\rm Edd.}(M_{\rm BH}) \right\} \,,
\ee
where $f_1^{\rm acc.} = 5.5\times 10^{-4}$ and $f_2^{\rm acc.} = 5\times 10^{-5}/{\rm Myr}$ is the fraction of the available gas that can be accreted by the SMBH, 
\be
    f_{\rm rem.}(M,z) \equiv 1-(f_{\rm ej.}(M,z)+f_*(M,z)) f_{\rm cold}(M)
\ee    
is the total fraction of gas left after star formation and ejection of gas by SNe, $f_{\rm Edd.}$ is the fraction of the Eddington rate at which the SMBHs can maximally accrete and the Eddington accretion rate is given by
\be \label{eq:edd}
    \dot{M}_{\rm Edd.}(M_{\rm BH}) = \frac{4\pi G M_{\rm BH} m_p}{\epsilon_r \sigma_T} \approx 2.2\times 10^{-3} m_{\rm BH}(M_{\rm BH})/{\rm Myr} \,,
\ee
where $m_p$ is the proton mass and $\sigma_T$ is the Thomson cross section. The physical interpretation is that in low-mass halos the SN feedback prevents efficient accretion, and in heavy halos, the accretion is limited by the Eddington rate and the amount of available gas.

We initiate the evolution by planting a seed of mass $m_{\rm seed}$ at some redshift $z_{\rm seed}$ in all halos that are heavier than some minimal mass $M_{\rm seed}$. Note that we evolve only the expected BH mass, $M_{\rm BH} = \int \td P(m_{\rm BH}) \,m_{\rm BH}$, and do not include the inevitable scatter in the present computations. We therefore estimate the occupation fractions with Dirac delta functions
\be
    \frac{\td P(m_{\rm BH}|M_*,z)}{\td m_{\rm BH}} = \frac{M_{\rm BH}(M_*,z)}{m_{\rm BH}} \delta\left[m_{\rm BH} - m_{\rm BH}(M_*,z)\right] \,,
\ee
where
\be
    m_{\rm BH}(M_*,z) = \max\left[m_{\rm seed},M_{\rm BH}(M_*,z)\right]
\ee
corresponds to the proper BH mass. This is the BH mass that enters the accretion rate in Eq.~\eqref{eq:edd} and that we compare with observations in the main text.

\begin{figure}
    \centering
    \includegraphics[width=0.24\textwidth]{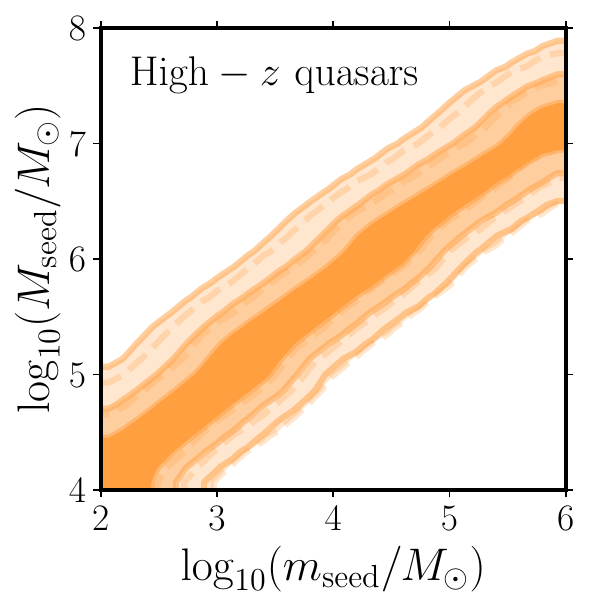}
    \includegraphics[width=0.24\textwidth]{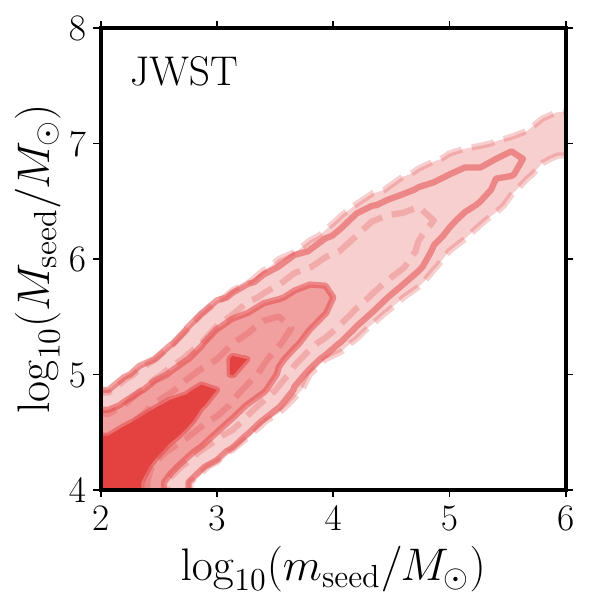} 
    \includegraphics[width=0.24\textwidth]{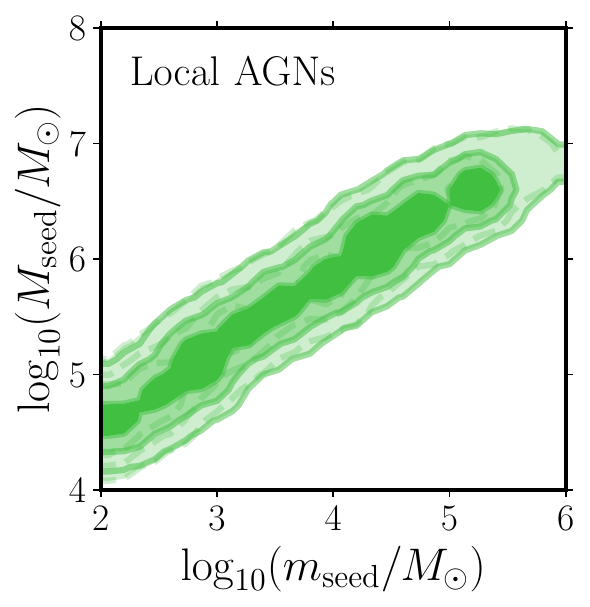}
    \includegraphics[width=0.24\textwidth]{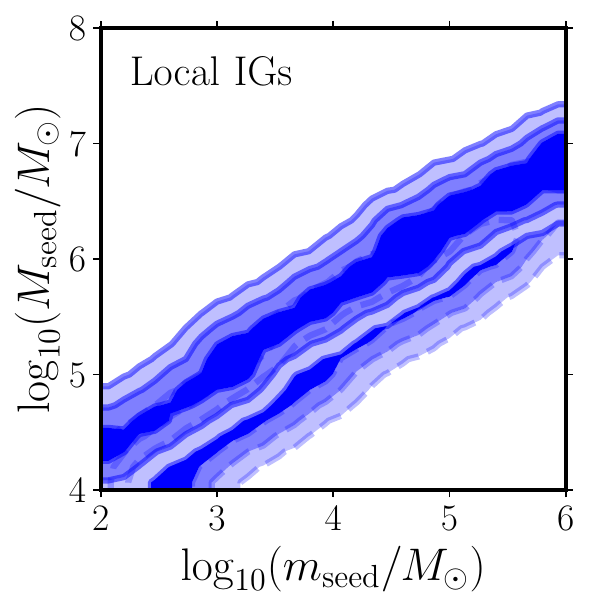}
    \caption{Fits to the stellar mass-BH mass data as functions of the seed BH mass $m_{\rm seed}$ and minimal halo mass $M_{\rm seed}$ where the seeds are inserted at $z_{\rm seed} = 20$, for $p_{\rm BH} = 1$ without spread in the predicted relation in solid and with a log-normal spread ($\sigma = 0.5$) in dashed.}
    \label{fig:fits_to_SMBH_data_0.5}
\end{figure}

\begin{figure}
    \centering
    \includegraphics[width=0.24\textwidth]{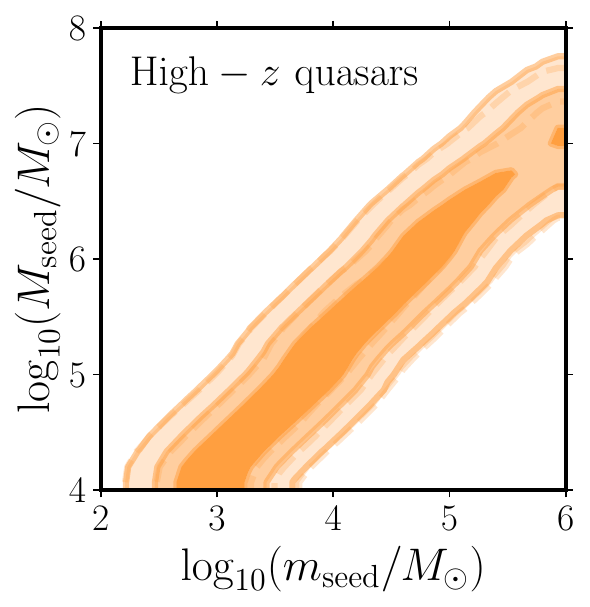}
    \includegraphics[width=0.24\textwidth]{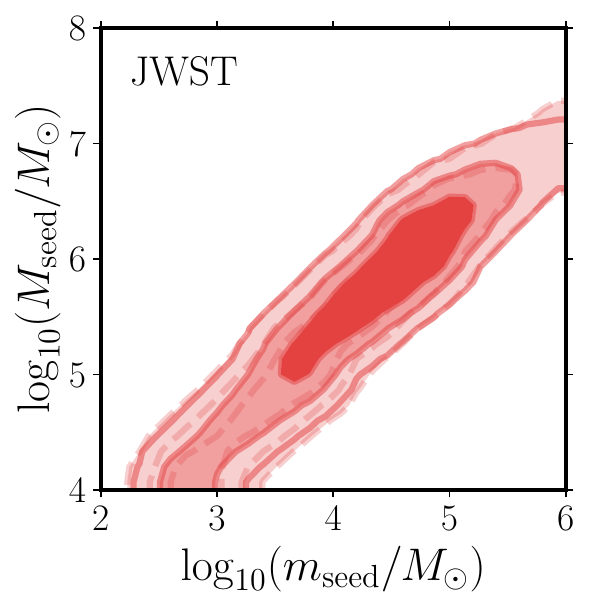}
    \includegraphics[width=0.24\textwidth]{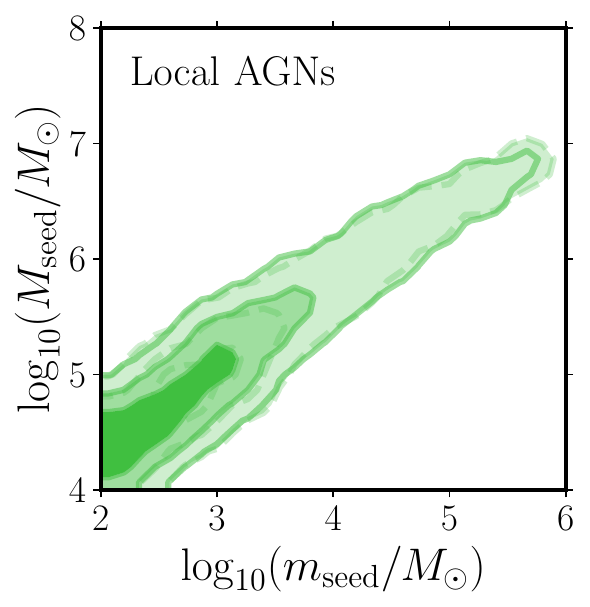}
    \includegraphics[width=0.24\textwidth]{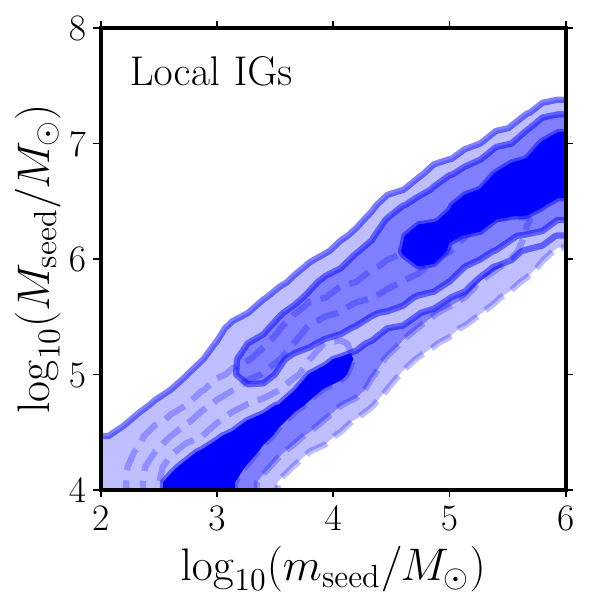}
    \caption{As in Fig.~\ref{fig:fits_to_SMBH_data_0.5} for $p_{\rm BH} = 0.1$.}
    \label{fig:fits_to_SMBH_data_0.1_0.5}
\end{figure}

As mentioned above, our model calculations are for means in the distribution of SMBH masses, and do not include the inevitable spread. Including this within the growth formalism we have described would cause the simple additive process to become a nonlinear convolution. In order to explore how a spread could change the results, we include ad-hoc spread in the distribution of BH masses only in the final comparison with the data, but the evolution is computed only using the mean of the distribution. We use the lognormal distribution
\be 
    \frac{\td P(m_{\rm BH}|M_*,z)}{\td \log_{10} \!m_{\rm BH}} =  \frac{M_{\rm BH}(M_*,z)}{m_{\rm BH}}\mathcal{N}(\log_{10} m_{\rm BH} |\log_{10} m_{\rm BH}(M_*,z),\sigma)
\ee
for this purpose, and assume $\sigma = 0.5$. Results from modifying the distribution in this way are shown in Fig.~\ref{fig:fits_to_SMBH_data_0.5} for $p_{\rm BH}=1$ and in Fig.~\ref{fig:fits_to_SMBH_data_0.1_0.5} for $p_{\rm BH}=0.1$ and included in Table~1, and are commented in the main text.

\begin{figure}
    \centering
    \includegraphics[width=0.24\textwidth]{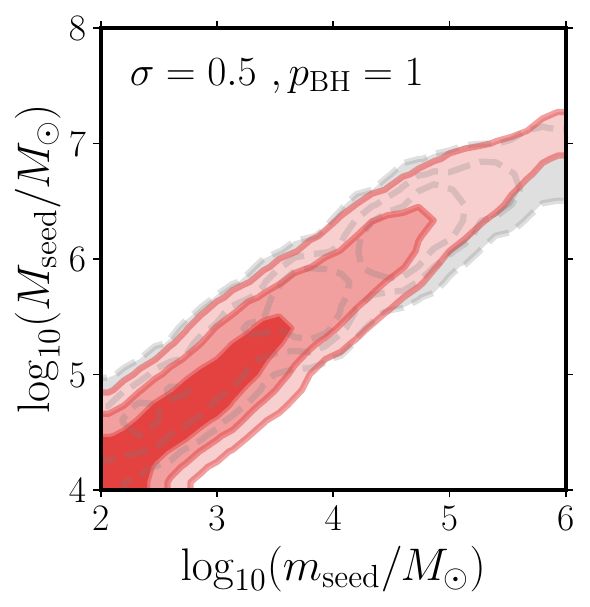} \hspace{5mm}
    \includegraphics[width=0.24\textwidth]{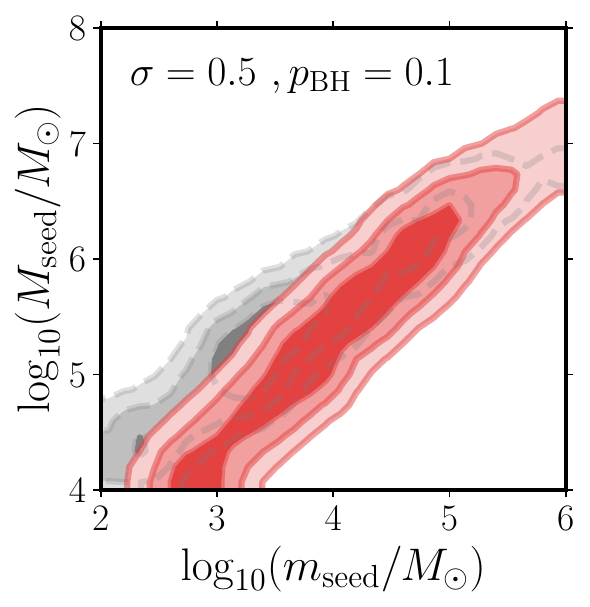}
    \caption{As in Fig.~\ref{fig:global_SMBH_data}, allowing for a lognormal spread in the model predictions with $\sigma=0.5$ for $p_{\rm BH} = 1$ in the left panel and for $p_{\rm BH} = 0.1$ in the right panel.}
    \label{fig:fits_to_SMBH_data_2}
\end{figure}

\end{document}